# Electron-Energy-Loss Spectra of Free-Standing Silicene

Luis M. Priede[1,a*] and Lilia Meza-Montes[2,b]

[1]Posgrado en Dispositivos Semiconductores, Benemérita Universidad Autónoma de Puebla, Puebla, Mexico

[2]Instituto de Física "Ing. Luis Rivera Terrazas", Benemérita Universidad Autónoma de Puebla, Puebla, Mexico

[a]luismpriede@gmail.com, blilia@ifuap.buap.mx

**Keywords:** silicene, EELS, tight binding, dielectric function.

**Abstract.** Silicene is becoming one of the most important two-dimensional materials. In this work, EEL Spectra were calculated for α-silicene (flat), and β-silicene (low-buckled, and theoretically the most stable). Band structures were determined using the semi-empirical Tight-Binding Method considering second nearest neighbors, $sp^3$ model, Harrison's rule, and Slater-Koster parameterization. The dielectric function was calculated within the Random Phase Approximation and a space discretization scheme. We found that, compared to bulk Si, additional resonances appear which are red-shifted. Buckling gives rise to a richer structure at low energy.

## Introduction

Silicene, which has a honeycomb structure with a two-atom basis (Fig. 1) is increasingly getting attention because it is a semimetal material with Dirac cones and thus, in principle, it has similar electronic properties to those of graphene [1-3]. However, unlike graphene, it has been predicted that the most stable structure has B atoms displaced $d = 0.44$ Å ($\theta = 101.18°$) in $z$-direction (downwards in our case, see Fig. 2) [3-5] called *β*-silicene. Free-standing *β*-silicene has lattice parameter of $a = 3.89$ Å and bond length of $a_e = 2.25$ Å [3, 6], which are larger compared with free-standing *α*-silicene, whose lattice parameter is $a = 3.86$ Å and bond length of $a_e = 2.228$ Å [2].

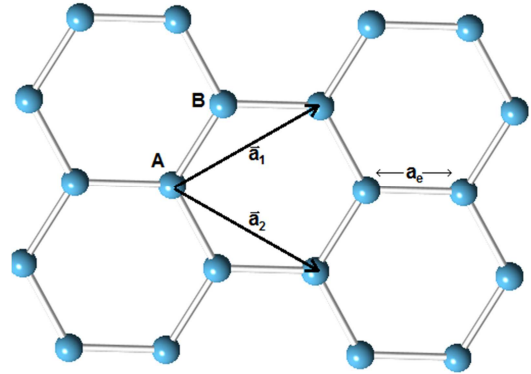

Fig. 1: Honeycomb structure.

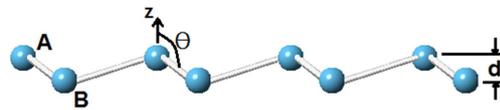

Fig. 2: Side view showing the buckled angle θ.

Silicene has been successfully synthesized on substrates, with different experimental procedures, so that the atomic structure depends on growth conditions. For example, it has been epitaxially deposited on Ag (111) [7] and Ir (111) [8] substrates, with B atoms displacement of 0.75 Å and 0.83 Å, respectively. On the other hand, Electron Energy Loss Spectroscopy (EELS) is a material characterization method used to study element composition, coordination, and electronic structure and it has been successfully applied to bidimensional materials [9]. High speed electrons are collimated towards the material. Some of the electrons will undergo inelastic scattering by the electrons in the material. The amount of lost energy can be measured via an electron spectrometer and interpreted in terms of the origin of the energy loss.

Here, we calculated EEL spectra in the low-energy regime for free-standing *α*-silicene (flat, $\theta = 90°$) and *β*-silicene (buckled, $\theta = 101.18°$) in the framework of the semi-empirical Tight-Binding

Method (TB) [10]. Band structures were determined considering second nearest neighbors, $sp^3$ model, Harrison's rule, and Slater-Koster parameterization [11-13].

**Theory**

We follow the work by Delerue *et al.* [10], who have developed a formalism to study silicon nanostructures within the TB framework. For systems with translational symmetry, we consider the atoms located at the sites $\mathbf{R}_{j,j_0}$ where $j_0$ refers to the atom in the $j$ cell. Thus, the one-electron Bloch wave functions $\psi_k(\mathbf{r})$ are expanded in the atomic basis $\phi_{j_0,\alpha}(\mathbf{r} - \mathbf{R}_{j,j_0})$ by

$$\psi_k(\mathbf{r}) = \frac{1}{\sqrt{N}} \sum_{j,j_0,\alpha} C_{j_0,\alpha} e^{i\mathbf{k}\cdot\mathbf{R}_{j,j_0}} \phi_{j_0,\alpha}(\mathbf{r} - \mathbf{R}_{j,j_0}), \tag{1}$$

where $\alpha$ refers to the atomic orbitals $s$, $p_x$, $p_y$, and $p_z$ while $N$ is the number of atoms in the system. Using this wave function, dielectric function is calculated within the Random Phase Approximation (RPA) [14], by means of Fourier transformation and discretization of real space, which allow a matrix formulation of dielectric function

$$\varepsilon(\omega, \mathbf{q}) = I - V(\omega, \mathbf{q})P(\omega, \mathbf{q}), \tag{2}$$

where $I$ represents the identity matrix (2×2 in this case), the polarization matrix is given by

$$P_{i_0,j_0} = 2\sum_k \sum_{l \in VB} \sum_{m \in CB} \left\{ \frac{\left[\sum_\alpha C_{i_0,\alpha}(l,\mathbf{k}+\mathbf{q}) C^*_{i_0,\alpha}(m,\mathbf{k})\right]\left[\sum_\alpha C^*_{j_0,\alpha}(l,\mathbf{k}+\mathbf{q}) C_{j_0,\alpha}(m,\mathbf{k})\right]}{\epsilon_l(\mathbf{k}+\mathbf{q}) - \epsilon_m(\mathbf{k}) - \omega - i\delta} \right.$$

$$\left. - \frac{\left[\sum_\alpha C^*_{i_0,\alpha}(l,\mathbf{k}-\mathbf{q}) C_{i_0,\alpha}(m,\mathbf{k})\right]\left[\sum_\alpha C_{j_0,\alpha}(l,\mathbf{k}-\mathbf{q}) C^*_{j_0,\alpha}(m,\mathbf{k})\right]}{\epsilon_m(\mathbf{k}) - \epsilon_l(\mathbf{k}-\mathbf{q}) - \omega - i\delta} \right\}, \tag{3}$$

and the Coulombic potential $V(\mathbf{q})$ used here is the one derived by Lannoo [15]. Monkhorst-Pack Method [16] is applied for the calculation of sums in the Irreducible Brillouin Zone (IBZ) (Fig. 3).

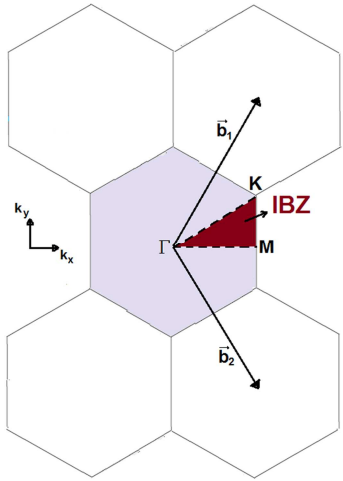

Fig. 3: Silicene reciprocal lattice.

The energy loss of fast electrons interacting with the material is related to $\varepsilon(\omega, \mathbf{q})$ and the response is usually described in terms of $-Im(\varepsilon^{-1}(\omega, \mathbf{q}))$. Therefore, results for this Electron Energy Loss Function (EELF) are presented. Here the longitudinal case is addressed, *i. e.*, it is considered that incident electrons move along the *x*-axis, see Ref. [10] for more details.

**Results**

Given the number of orbitals included in the basis, band structures of both α- and β-silicene (Fig. 4(a) and 4(b), respectively) have 8 energy bands (4 conduction band, and 4 valence band). The Fermi level is not at zero, due to the Slater-Koster parameters used here are for bulk silicon [11].

The band structures show the graphene-like Dirac cones at K points. However, in contrast to graphene, they are asymmetric around the Fermi level, as it can be noticed at DOS charts.

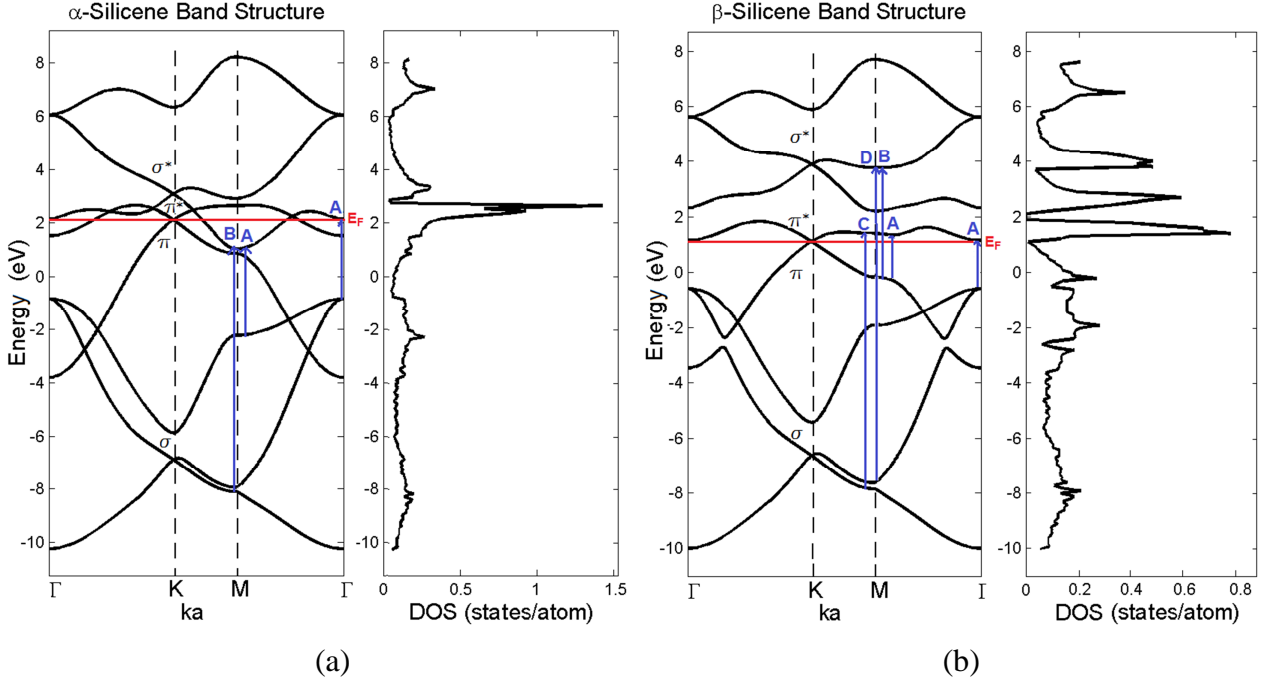

(a) (b)

Fig. 4: Band Structure and density of states DOS of (a) α-silicene and (b) β-silicene. Some electronic transitions are indicated (see Fig. 5 and 7 below).

Our results succesfully reproduce the band structure of Ref. [11] for α-silicene. As $\theta$ increases to go from α-silicene to β-silicene, gaps appear and bands become flat. Qualitative agreement with previous *ab initio* calculations is found [3, 20], differences at high-energy bands appear as expected.

Band structure of α-silicene has the first conduction band below the Fermi level, and therefore the DOS at this energy has a finite value, indicating that it is a metal; however, it is not the most stable theoretically.

On the other hand, β-silicene band structure shows that buckling introduces flattening and gaps in the band structure, changing in turn its DOS. In this case, all of the conduction bands are above the Fermi Level. Therefore, it is a semimetal material with zero bandgap, such as graphene [22].

Dielectric Function and corresponding EELF of α-silicene when $q\rightarrow 0$ are shown in Fig. 5. The static dielectric constant ($\varepsilon_{st}$) is 27, and the principal peaks in EELF are at A = 3 and B = 9.5, in eV. Electronic transitions potentially leading to these peaks are shown in Fig. 4(a), where the peak A is due to the transitions between the 2$^{nd}$ valence band, and the 2$^{nd}$ conduction band around point M and point Γ. Peak B arises from the transitions between the 4$^{th}$ valence band, and the 2$^{nd}$ conduction band along the K-M direction (also the small peak at ~11 eV).

Dielectric function and EELF calculated for the α-silicene when $\mathbf{q}a=(\pi/5, 0)$ are shown in Fig. 6. The static dielectric constant decreases to 16, and peaks around the same position as before, A = 3.2 eV and B = 8.5 eV. Also noticiable are the change in intensity and additional peaks at 8.5 and 11.3 eV, a consequence of interference of the terms in Eq. (3).

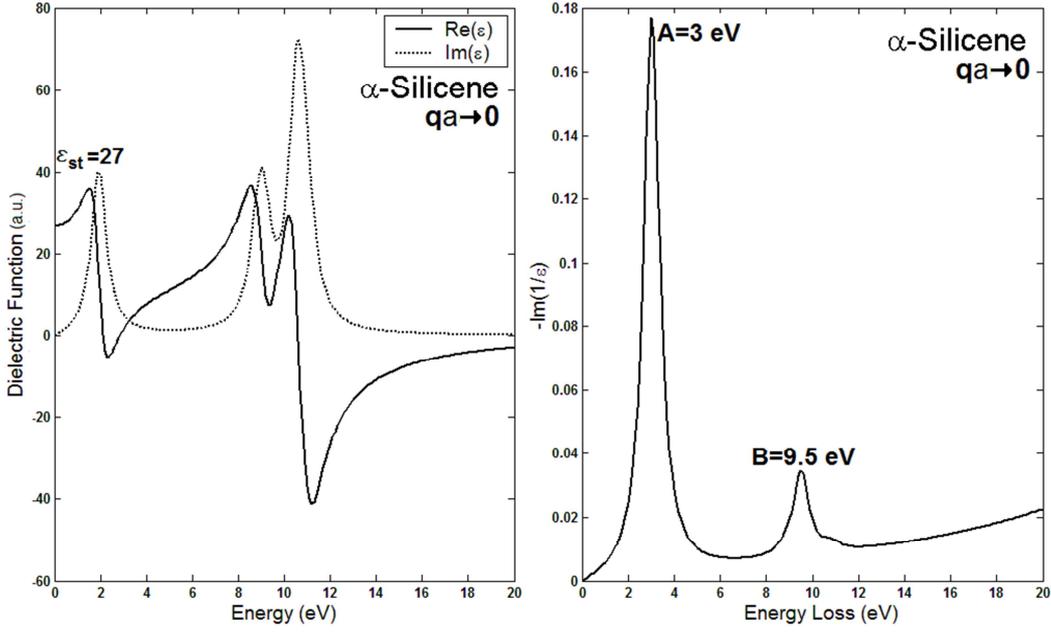

Fig. 5: Dielectric Function and Electron Energy Loss Function of α-silicene for **q**→0.

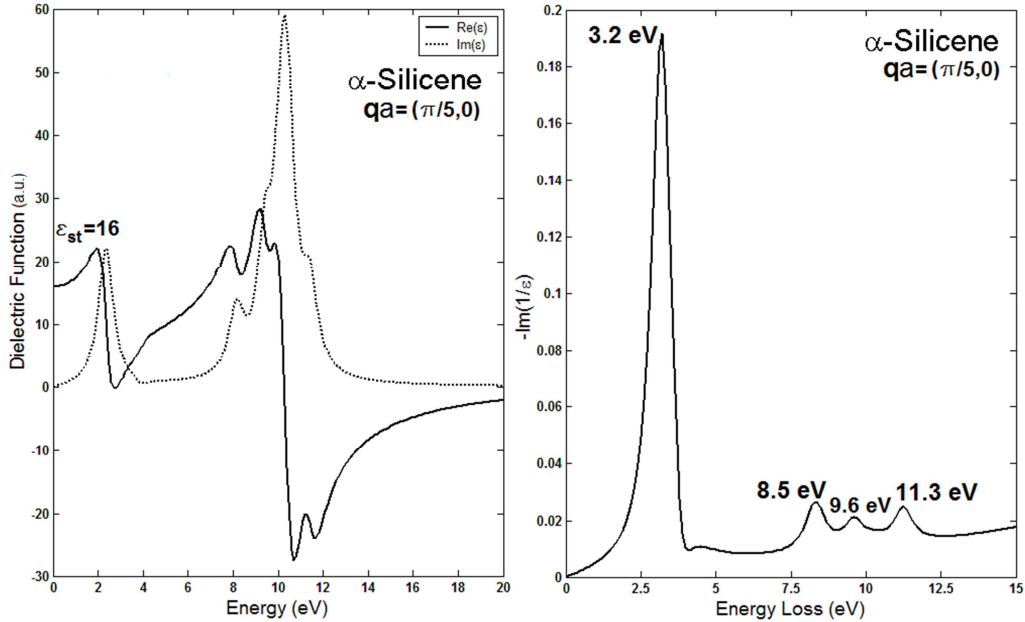

Fig. 6: Dielectric Function and Electron Energy Loss Function of α-silicene for **q**a=(π/5, 0).

As for the case of *β*-silicene, the Dielectric Function and EELF for q→0, show in Fig. 7 that the static dielectric constant is 16.4 and that new peaks appear at low energies compared to the flat case, due to separation and flattening of the energy bands.

The peaks calculated from its dielectric function are (in eV) at: A = 2, B = 4.5, C = 9.1, and D = 11. Electronic transitions that might occur are depicted in Fig.4(b). Peak A is caused by the transition between π and π* bands around the point M and Γ, B takes place between the π band and the $3^{th}$ conduction band in the M-Γ direction, C between the $4^{th}$ valence band and the π* conduction band, and D peak between the third valence and conduction bands. K. Chinnathambi *et al.* [20] reported *ab initio* absorption spectra, related to the dielectric function, with a peak at 1.74 eV and one more intense centered at 3.94 eV which approach our A and B calculated peaks.

Finally, we present the dielectric function and EELF for **q**a=(π/5, 0) (Fig. 8). The static dielectric constant decreases to 12.9, the two peaks A and B are kept at the same energy. As in the previous case, there are three more peaks at 8.5, 10.3 and 11.8, all in eV. The static dielectric constant diminishes as q decreases, and compared to the flat case, $\varepsilon_{st}$ for *β*-silicene is smaller for the chosen **q**'s. EEL spectra show similar structure at high energy for $\mathbf{q}a = (\pi/5, 0)$ in both cases, due to the similarity between DOS. However, around the Fermi energy, in the case of *β*-silicene flattening of bands introduces a richer structure of DOS giving rise to additional peaks.

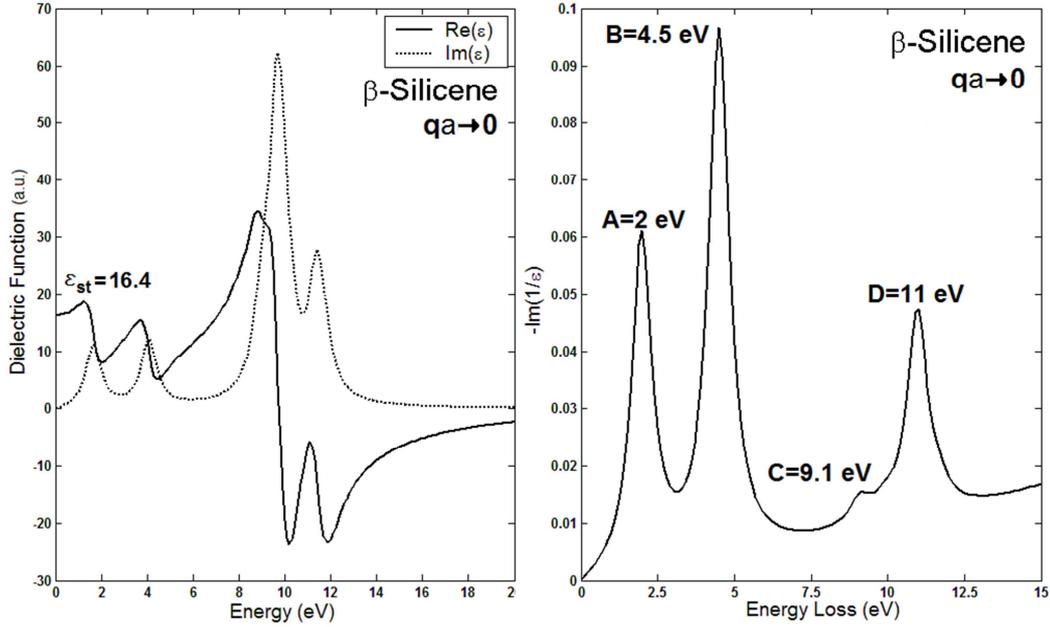

Fig. 7: Dielectric Function and Electron Energy Loss Function of *β*-silicene for **q**→0.

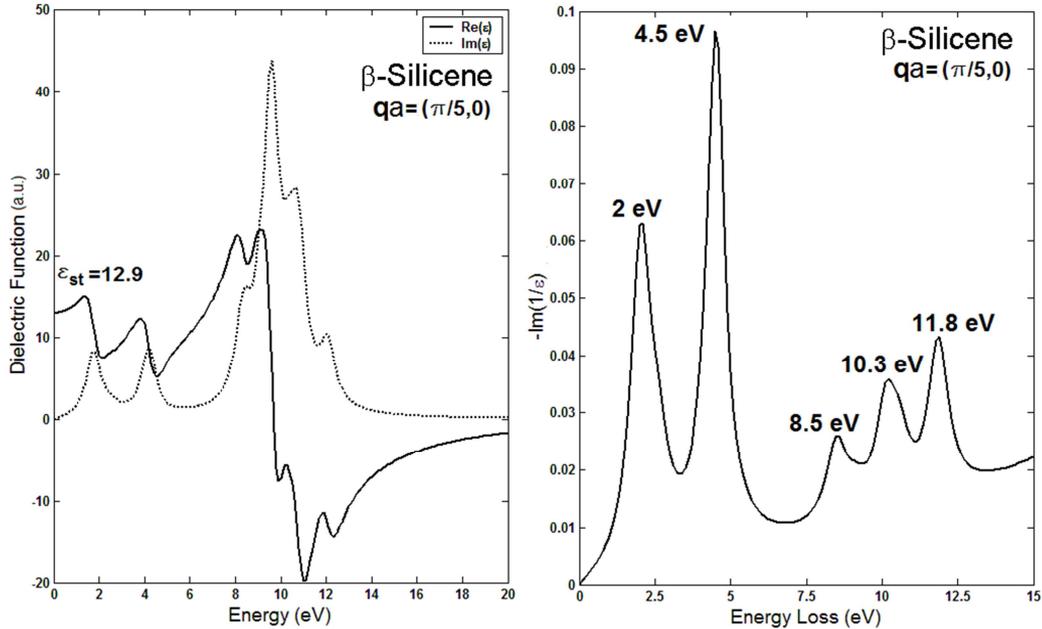

Fig. 8: Dielectric Function and Electron Energy Loss Function of *β*-silicene for $\mathbf{q}a=(\pi/5, 0)$.

**Summary**

Buckling introduces flattening and gaps in the band structure, which in turn modifies the DOS around the Fermi energy. These changes introduce additional peaks in the dielectric function at low energies while for high energies only peak intensity changes. As for the energy loss spectra, similar changes are observed, *i. e.*, buckling gives rise to additional resonances at low energy and changes only in intensity at high energy. Compared to bulk silicon, for which only one peak appears at approximately 16.9 eV, resonances in silicene are red-shifted and, as mentioned above, additional peaks appear. Our results are consistent with DFT calculations.

**Acknowledgements**

This work was partially supported by CONACyT CB/2009/133516 and VIEP-BUAP. We thank E.Gómez-Barojas for useful discussions.